\documentclass{aa}  

\usepackage{graphicx}
\usepackage{ulem}
\usepackage{epstopdf}
\usepackage{multirow}
\usepackage{array,makecell}
\setcellgapes{1pt}
\makegapedcells
\usepackage{txfonts}
\begin{document}

   \title{Lunar laser ranging in infrared at the Grasse laser station}

   \subtitle{}

   \author{C. Courde\inst{1}\and J.M. Torre\inst{1}\and E. Samain\inst{1}\and G. Martinot-Lagarde\inst{1}\and M. Aimar\inst{1}\and D. Albanese\inst{1}\and P. Exertier\inst{1}\and A. Fienga\inst{1}\and H. Mariey\inst{1}\and G. Metris\inst{1}\and H. Viot\inst{1}\and V. Viswanathan\inst{1}}

   \institute{Université Côte d'Azur, CNRS, Observatoire de la Côte d'Azur, IRD, Géoazur UMR 7329, 2130 route de l’Observatoire, 06460 Caussols, France\\
              \email{clement.courde@geoazur.unice.fr}
             }

   \date{Received March 24, 2016; accepted }

% \abstract{}{}{}{}{} 
% 5 {} token are mandatory

  \abstract{For many years, lunar laser ranging (LLR) observations using a green wavelength have suffered an inhomogeneity problem both temporally and spatially. This paper reports on the implementation of a new infrared detection at the Grasse LLR station and describes how infrared telemetry improves this situation. Our first results show that infrared detection permits us to densify the observations and allows measurements during the new and the full Moon periods. The link budget improvement leads to homogeneous telemetric measurements on each lunar retro-reflector. Finally, a surprising result is obtained on the Lunokhod 2 array which attains the same efficiency as Lunokhod 1 with an infrared laser link, although those two targets exhibit a differential efficiency of six with a green laser link.}
   %{For many years, Lunar Laser Ranging observations using a green wavelength, suffer of an inhomogeneity problem both temporally and spatially. }
   %{This paper reports on the implementation of a new infrared detection at the Grasse LLR station and describes how infrared telemetry improves this situation. }
   %{Our first results show that infrared detection permits to densify the observations and allows measurements during the new and the full Moon periods. The link budget improvement leads to homogeneous telemetric measurements on each lunar retro-reflector. Finally, a surprising result is obtained on the Lunokhod 2 array which procures the same efficiency than Lunokhod 1 with an infrared laser link, although those two targets exhibit a differential efficiency of 6 with a green laser link.}


   \maketitle

%
%-------------------------------------------------------------------

\section{Introduction}

  The Lunar laser ranging (LLR) technique has existed since the beginning of the Moon conquest in 1969. Thanks to the installation of retro-reflectors on the Moon surface, it was then possible to measure the Earth-Moon distance. LLR exploits two kinds of retro-reflector arrays: three American arrays (A11, A14 $\&$ A15) placed during the Apollo missions 11, 14 $\&$ 15 ;  and two French arrays (L1 $\&$ L2) loaded on Soviet rovers Lunokhod 1 $\&$ 2. 
  
  Long-term observations of Earth-Moon distances, associated with a continuous improvement of the metrological performances, contributed significantly to the understanding of the Moon internal structure, its dynamics, and its interactions with the Earth. The Earth-Moon system is an essential laboratory for the study of the solar system and in particular its formation mechanisms. Despite the analysis of data collected by the Apollo missions, space missions in orbit (i.e. the GRAIL mission) and 45 years of LLR data, many questions remain, including on the internal physics of the Moon. Although the Moon formation scenario seems almost established, the presence of a fluid without a solid core could jeopardise the mechanisms of differentiation used until now in scenarios of formation and evolution of the solar system \citep{Williams2012}. Moreover, LLR data are essential to fundamental physics tests \citep{Williams2012,Fienga2014} that require regular data sampling spread out over several dozens of years. Lunar laser ranging data are currently providing on-ground tests of the equivalence principle (EP) that are as accurate as those provided by torsion balances \citep{Williams2012}. Because the Earth and Moon are large bodies, LLR measurements also have the advantage of allowing us to test the contribution of the internal energy to the EP. The equivalence principle is at the heart of general relativity and many alternative theories of gravity predict its violation. A complete review of research undertaken using LLR data can be found in \citep{Murphy2013,Merkowitz2010,Muller2012}. 
  
  Since its beginning, the LLR technique has been based on the time-of-flight measurement of an optical pulse during its round-trip between the Earth and the Moon. The LLR technique is similar to satellite laser ranging (SLR). However, due to the large distance between the Earth and the Moon, the performance of the sub-systems are quite different. The round trip loss, which is the ratio of photon numbers received and sent, is around 1:10$^{18}$, meaning that large and costly SLR facilities are required to compensate for this. Among the 40 SLR stations of the International Laser Ranging Service (ILRS) network, only a few stations are able to detect echoes on the Moon. Consequently, only four stations supply Earth-Moon range observations to the ILRS database: APOLLO (New Mexico, USA), McDonald (Texas, USA), Matera (Italy) and Grasse (France). Apart from APOLLO, the other LLR stations suffer from a low detection rate, for example, the photon flux at Grasse on A15 is roughly 0.1 photon.s$^{-1}$ compared to the 63 photons.s$^{-1}$ at APOLLO \citep{Murphy2013}. The quality of LLR data is quite different between stations and only APOLLO, thanks to its greater detection rate, has measurements limited by the orientation of the reflector array and the associated spread of pulse return times. 
  The single-photon events dated by LLR stations over observations of ten minutes are combined to generate “normal points” (NPs). These condensed range observations and the different conditions and corrections associated with them are formated in “consolidated range data” (CRD) files, that are then delivered to, and made available to the scientific community, by the ILRS.  These NPs are then compared to theories based on models of the solar system that include prescriptions of all the physical effects that have an impact on the measurement.  Several LLR models are under development around the world, at the following locations: the Jet Propulsion Laboratory (JPL), USA; the Harvard-Smithsonian Center for Astrophysics (CfA), USA; the Leibniz University in Hanover, Germany; the Paris Observatory, France; the Institute of Applied Astronomy of the Russian Academy of Sciences, Russia. In our laboratory, a team works on LLR reduction models using the GINS software (Geodesy by Simultaneous Digital Integration developed by the Centre National d'Etudes Spatiales (CNES)) \citep{marty2011gins}, as part of the LLR stepwise comparison study mentioned in \citep{Murphy2013}. The JPL model currently demonstrates  the best performance because weighted root-mean-square (RMS) residuals around 18 mm are obtained from the comparison with LLR NPs produced at Grasse \citep{vishnu} and at APOLLO \citep{Murphy2013}. However, APOLLO’s millimetric level performance \citep{Battat2009,Murphy2012} is ten times better than this difference, suggesting that the models need major improvements. In parallel with model improvements, investigations have to be undertaken into LLR data production. 	
  
  Despite around 40 years of continuous developments, acquisition of LLR data remains complicated. The distribution of LLR observations is not temporally and spatially homogenous. On the Earth, active LLR stations are only present in the northern hemisphere. This results in a non-complete coverage of the whole lunar orbit. It has also been demonstrated that an improvement of the temporal distribution could drive an improvement of the EP test, thanks to the solving of an asymmetry problem of the post-fit signal resulting from a cos D nodal signal in the residuals \citep{Nordtvedt,Muller1998}. 
  
  %----------------------------------------------------------------- one column figure
     \begin{figure}
     \centering
     \includegraphics[width=\hsize]{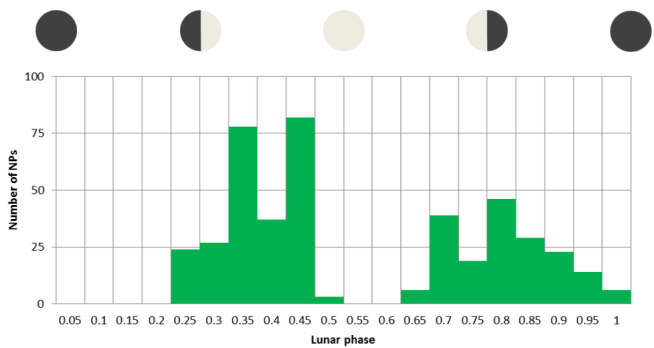}
        \caption{Number of NPs in 2014, from Grasse only, versus the lunar phase. All the observations were with the green wavelength. New Moon is depicted by a dark disk and matches lunar phase 0.0. Full Moon is depicted by a bright disk and matches lunar phase 0.5.}\label{fig:1}
     \end{figure}
  %-----------------------------------------------------------------
  
  %-------------------------------------- Two column figure (place early!)
    % \begin{figure*}
    % \centering
     %%%\includegraphics{empty.eps}
     %%%\includegraphics{empty.eps}
     % % %\includegraphics{img1.eps}
    % \caption{Number of NPs in 2014 versus the lunar phase. All the observations have
     %  been done with the green wavelength.}\label{fig:1}
      %\end{figure*}
  %
  
  In Fig. \ref{fig:1}, we show the NPs production of the Grasse station during the year 2014 versus the lunar phases.  During two particular phases, we see a lack of data. The first point is at the new Moon period, during which the Moon is not lit by the sun. Two difficulties must be overcome in this configuration. Firstly, we can no longer accurately correct the pointing by using our knowledge of the Moon topography near the target despite a good pointing model of the telescope. Secondly, we have to observe near the Sun, often at an angle less than 10$^\circ$ to it (the sun avoidance angle at Grasse is 6$^\circ$). This yields a high level of noise hiding many laser echoes. The presence of the Sun close to the Moon also makes it difficult to observe reference stars that are very useful to improve the pointing model of the telescope. The second point is at the full Moon, the period during which the Moon is totally lit by the Sun, this light strongly reduces the signal to noise ratio. The number of targets is also restricted to  the only reflectors not distorted too much by solar heating, namely A11, A14 $\&$ A15. 
  
\begin{figure}
       \centering
       \includegraphics[width=\hsize]{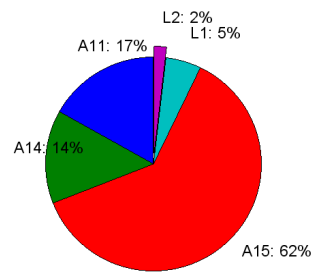}
          \caption{Distribution of the NP on the different lunar retro-reflectors during the year 2014 with green beam at the Grasse LLR station}\label{fig:2}
       \end{figure}

  Moreover, the daylight background noise often exceeds the return signal for green LLR observations. This introduces daily effects. An inhomogeneity of the retro-reflector tracking is present in addition to this temporal inhomogeneity, resulting in 60\% of the LLR NPs being obtained on A15 (see Fig. \ref{fig:2}). This observation is not particular to the Grasse facility but is shared in a similar way by the other LLR stations.  The predominance of A15 data is not ideal for lunar libration measurements. If APOLLO is able to operate during the full Moon, a reduction of the signal strength by a factor of ten around this period is observed \citep{murphy2010long,murphy2014lunar}. The authors presenting this observation proposed the presence of dust on the front surface of the corner-cube prisms as a possible explanation for the degradation of their reflectivity. The APOLLO data production is also not constant due to time-sharing with other programs and there is a gap in data at the new Moon. Considered together, these facts explain the inhomogeneity of LLR observations.  
  
   %----------------------------------------------------------------- one column figure
    %-----------------------------------------------------------------
  
  Background noise is mainly due to solar photons. The solar irradiance in infrared (IR) is approximately equal to half the irradiance in green \citep{Zissis1993}. This fact has led us to investigate whether the addition of IR detection in LLR observations can decrease these observation inhomogeneities. Lunar laser ranging in IR has already been performed but only at the Grasse laser station \citep{Veillet1989,Mangin1989}.Although promising results were obtained concerning the number of echoes, the detector performances in IR concerning the noise and the measurement precisions, were not suitable. In the following section, we briefly present the current status of the LLR ground station at Grasse and we report on the implementation of an InGaAs Single-Photon Avalanche Diode (SPAD).  In Sect. 3, we present our new results obtained in IR and the comparison with those in green. Finally, in Sect. 4, we present the scientific impact resulting from this change of wavelength.

%--------------------------------------------------------------------
\section{Lunar laser ranging ground station at Grasse}

\begin{figure}
         \centering
         \includegraphics[width=\hsize]{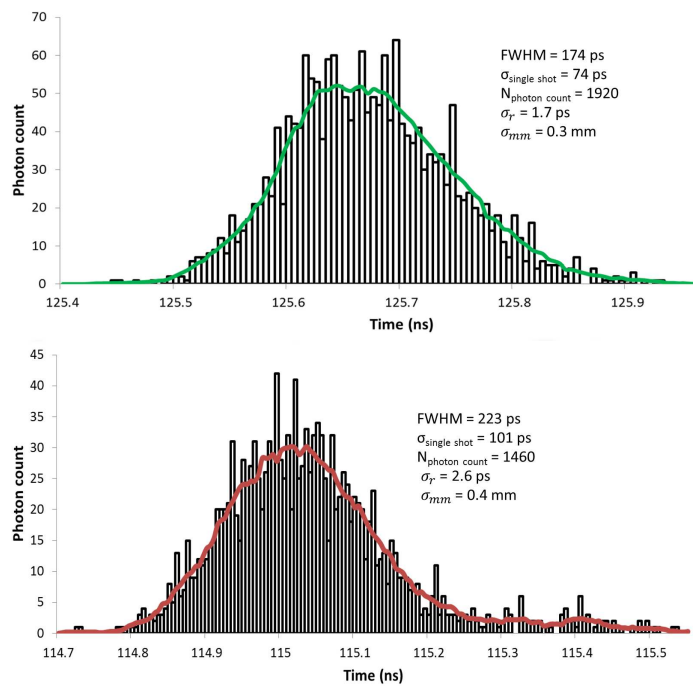}
            \caption{Histogram of the calibrations of the green and infrared detection channels. The histograms are created using a 5 ps binning. The traces in bold are a five-point moving average.}\label{fig:3}
         \end{figure}

%-------------------------------------- Two column figure (place early!)
  In France, the first echoes on the lunar reflectors were obtained in December 1970. A ruby laser was set on a telescope in the Pic du Midi Observatory. Due to some difficulties at Pic du Midi, a decision was taken to build an instrument dedicated to lunar laser telemetry based on a 1.5m telescope and to install it in the centre for geodynamic and astronomic studies (CERGA) (since then merged with the Nice Observatory to form the Côte d'Azur Observatory). Continuous observations were made from 1984 to 2005. The station exploited a Ruby laser operating at 694 nm until 1986, after which a Nd:YAG laser producing laser pulses at the 1064 nm wavelength was implemented. After the second harmonic generation, this laser operated at 532 nm because the green detector was the best. Since 1994 and the use of a SPAD, the measurement precision has reached the millimetric level \citep{Samain1998}. However, despite a high precision level only a small percent of LLR measurements have been obtained at the intrinsic precision of the Grasse LLR station mainly because of measurement dispersion coming from the orientation of the arrays \citep{Samain1998}. In addition, more than 60\% of LLR NPs have been obtained on A15, the largest reflector. Some tests have also been performed in IR with silicium photodiodes \citep{Samain1994,Schreiber1994}, but at the 1994 time, the precision level of IR detection was clearly insufficient. They had also a high level of internal noise. Thus, measurement precision was limited by the detector and not by the array orientation. Moreover, IR detectors based on InGaAs or Ge technologies were very noisy compared to green detectors and required complicated cooling systems \citep{Cova1994,prochazka1996large}.  
  Recent progress in InGaAs technology has changed this context \citep{Itzler2011}.  An InGaAs/InP single-photon avalanche diode (SPAD) from Princeton Lightwave has been purchased for the Grasse station. The model is a PGA-284 mounted inside a PGA-200 housing to take advantage of the three Peltier cooling stages. The quantum efficiency is about 20\% in Geiger mode and the active area has a diameter of 80 $\mu$m. The SPAD is used in gated-mode with active quenching. Characterization in the laboratory shows a dark-count rate of 28 kHz at -40$^\circ$C with +10V applied above the breakdown voltage. The timing jitter measured in the same conditions is about 109 ps full width at half maximum (FWHM) with the use of a Dassault event timer triggered at -100 mV. This level of precision is suitable for LLR. Finally, the time-walk, which represents a SPAD bias regarding the number of photons received, is about 100 ps.decade$^{-1}$. The laser can be configured to generate a beacon based on a single IR wavelength or double IR $\&$ green wavelengths. We can easily measure the distances with either the two wavelengths or only in IR. Two detection channels are placed in a box at the Nasmyth table. A dichroic mirror splits the green and the IR photons of the returning beam. The spectral filtering on each channel is achieved by two filters placed in series in front of each SPAD, namely an interference filter of 6 nm FWHM bandwidth and a Fabry-Perot filter of 0.12 nm FWHM bandwidth. Calibrations of both channels are achieved thanks to a corner cube placed at the telescope output. In these conditions, the contribution of all the sub-systems of the instrument are taken into account. The width of the returns from this corner cube give the single-shot precision of the station. The center of the returns gives both the delay relating to the total delays in the instrument and two times the distance between the corner cube and the intersection of the telescope axes. Figure \ref{fig:3} shows the histogram of the calibrations in green and in infrared. The precision of the calibration is 74 ps RMS for the green channel and 101 ps RMS for the IR channel.  
   %----------------------------------------------------------------- one column figure
      %-----------------------------------------------------------------

 \section{Results}
 \subsection{Comparison of the different lunar retro-reflectors in infrared}
 \subsubsection{Characteristics of the lunar retro-reflectors}
 Each array is an assembly of individual corner cube retro-reflectors. For the Apollo arrays, each corner cube is uncoated and operated via total internal reflections \citep{Chang1972}. The advantage of this approach is to offer a total reflection of the incident light. As a result, the light absorption is prevented, guaranteeing smaller thermal gradients within the corner cube when they are illuminated by the sunlight. The corner cubes are circular with a 38 mm diameter. The arrays A11 $\&$ A14 are identical squares of size 46cm x 46cm. The 100 corner cubes are distributed on 10 rows and 10 columns. The Apollo 15 array is composed of 300 corner cube retro-reflectors shared on two panels: one with 204 corner cube reflectors  distributed on 17 columns and 12 rows ; and a second with 96 corner cube reflectors distributed on 8 columns and 12 rows. The total array dimensions are 1.05 m x 0.64 m. 
 The Lunokhod retro-reflectors have been purchased by CNES \citep{Fournet1972} and are identical for the two rovers. The arrays are composed of 14 triangular corner cubes, nearly-joined and arranged on two lines, with external dimensions of 44cm x 19cm. These corner cubes are quite different from the Apollo corner cubes because they are composed of rear-reflecting silica cubes that have a triangular edge length of 10.6 cm and a silver coating on their rear faces. The corners of the triangle have been depressed so that the efficient part of the corner cube corresponds roughly to a regular hexagon with a side of 35.5mm. The Lunokhod arrays have been optimised for observations at the wavelength of a Ruby laser. The array design has been based on the assumption of the isothermal conditions. The goal of lunar day operation has been discarded because the arrays are unusable when they are lit by the sun at normal incidence. The times of day during which the array is not degraded (sunrise and sunset) were determined through thermal testing. 
 %
  %-------------------------------------------------------------
  %                                             Simple A&A Table
  %-------------------------------------------------------------
  %
  \begin{table*}
  \caption{Efficiency reduction factor ($\eta$) of one Lunokhod corner cube at different solar angles \citep{Fournet1972}}             % title of Table
  \label{table:1}      % is used to refer this table in the text
  \centering                          % used for centering table
  \begin{tabular}{|c|c|c|c|c|c|c|}        % centered columns (4 columns)
  \hline                 % inserts double horizontal lines
  Sun elevation & night & 90$^\circ{}$ & 85$^\circ{}$ & 71$^\circ{}$ & 42$^\circ{}$ & 23$^\circ{}$ \\    % table heading 
  (zenith angle in $^\circ{}$) & & & & & & \\
  \hline                        % inserts single horizontal line
     $\eta$ & 0.82 & 0.74 & 0.44 & 0.16 & 0.09 & 0.045 \\      % inserting body of the table
  \hline                                   %inserts single line
  \end{tabular}
  \end{table*}
 
 Table \ref{table:1} summarises the efficiency reduction factor of one corner cube for each solar incident angle. The array efficiency stays above its maximal value during the twelve hours that follow the sunrise \citep{Fournet1972}.

\subsubsection{Theoretical estimations}
Using the same analysis as in \citep{Murphy2011}, we can estimate each reflector's performance for the wavelength of 1064 nm. At normal incidence, the hexagonal reflecting area of a Lunokhod corner cube is equal to 3274 mm$^2$ and the Apollo corner cube has a reflective area of 1134 mm$^2$.The far-field diffraction peak intensity of each corner cube scales with the square of the reflecting area. Thus, each Apollo corner cube dilutes the retro-reflected light eight times more than each Lunokhod corner cube. The central intensity of the uncoated Apollo corner cube is only 27\% of a perfect Airy pattern arising from a perfectly coated corner cube. Therefore, by taking into account the number of corner cubes and the central intensity of each corner cube, we obtain that the central intensity of the Lunokhod array's reflected beam should be 3.5 times stronger than that of the A11 $\&$ A14 arrays returns and 1.2 times stronger than that of the A15 return. However, we have to take into account an average velocity aberration of 5 $\mu$rad due to the transverse motion of the Moon together with Earth rotation. The line of sight of the telescope is not perfectly aligned with the central intensity of the far-field diffraction pattern, yielding not only an attenuation which is more important for Lunokhod than Apollo corner cubes (due to the narrower diffraction pattern) but also an attenuation which will depend on the wavelength. The intensity collected by the telescope is given by the Airy pattern at an angle $\delta$ from the centre. For a corner cube with a circular aperture and an effective diameter $D_{cc}$ this intensity is equal to \begin{equation} 4\frac{J_1^2(x)}{x^2}, \end{equation}
where $x=\frac{\pi{}\delta{}D_{cc}}{\lambda{}}$and $J_1$ is the Bessel function of the first order.
$D_{cc}$ is equal to 3.8 cm for Apollo and 6.5 cm for Lunokhod. Consequently, the signal due to an average velocity aberration of 5 $\mu$rad is 0.79 and 0.92 times the central intensity of the far-field diffraction pattern, respectively for the Lunokhod and Apollo corner cubes. Finally, by taking into account the efficiency of the arrays given above, we obtain that in IR the central intensity of the Lunokhod arrays should be 3 times stronger than the A11 or A14 arrays and should be equal to the A15 array. 

 \begin{eqnarray}
 \frac{central\,intensity\,L1\,(or\,L2)\,array}{central\,intensity\,A11\,(or\,A14)\,array} = \nonumber \\ (\frac{L1\,reflecting\,area}{A11\,reflecting\,area})^2 \cdot (\frac{L1\,corner\,cube\,efficiency}{A11\,corner\,cube\,efficiency})  \cdot \nonumber \\
 (\frac{L1\,corner\,cube\,number}{A11\,corner\,cube\,number}) \cdot (\frac{L1\,velocity\,aberration\,effect}{A11\,velocity\,aberration\,effect})  \nonumber \\ 
  = (\frac{3274}{1134})^2 \cdot (\frac{0.82}{0.27}) \cdot (\frac{14}{100}) \cdot (\frac{0.79}{0.92}) \nonumber \\
   \approx 3, \nonumber \\
 \\[0.1cm]
 \frac{central\,intensity\,L1\,(or\,L2)\,array}{central\,intensity\,A15\,array} = \nonumber \\
 (\frac{L1\,reflecting\,area}{A11\,reflecting\,area})^2 \cdot (\frac{L1\,corner\,cube\,efficiency}{A11\,corner\,cube\,efficiency}) \cdot \nonumber \\
 (\frac{L1\,corner\,cube\,number}{A11\,corner\,cube\,number}) \cdot (\frac{L1\,velocity\,aberration\,effect}{A11\,velocity\,aberration\,effect}) \nonumber \\ 
   = (\frac{3274}{1134})^2 \cdot (\frac{0.82}{0.27}) \cdot (\frac{14}{300}) \cdot (\frac{0.79}{0.92}) \nonumber \\
   \approx 1  \nonumber \\
 \nonumber 
 \end{eqnarray}

 \subsubsection{Measurements}
 We compared the signal strength of the different lunar retro-reflectors in infrared. These measurements are quite difficult due to the strong variations of the photon flux with the atmosphere, the stability of the telescope pointing and the elevation angle. During specific campaigns, we measured the temporal photon flux. Each retro-reflector was followed for 5 minutes in order not to be disturbed by changes in observation conditions. Table \ref{table:2} gives an example of data obtained during one night near the last quarter (lunar phase of 0.79), and the corresponding photon flux ratios. 
 
 %-------------------------------------------------------------
 %                                             Simple A&A Table
 %-------------------------------------------------------------
 %
 \begin{table*}
 \caption{Measurements of the photon flux ratio between A11 \& L2 in IR.}             % title of Table
 \label{table:2}      % is used to refer this table in the text
 \centering                          % used for centering table
 \begin{tabular}{|c|c|c|}        % centered columns (4 columns)
   \hline
 \multicolumn{3}{|c|}{03/12/2015: Sun elevation 79.7$^\circ{}$ ; lunar phase 0.793} \\
 \hline
  Array & A11 & L2 \\
 \hline                                   %inserts single line
 Observation conditions & dark & dark \\
 \hline
 Nb of measurement series & 13 & 13 \\
 \hline
 Cumulative photon number & 539 & 1533 \\
 \hline
 Cumulative observation duration (s) & 3999 & 3545 \\
 \hline
 Photon flux mean (photon.s$^{-1}$) & 0.14 & 0.4 \\
 \hline
 Photon flux standard deviation (photon.s$^{-1}$) & 0.06 & 0.1 \\
 \hline
 Photon flux max (photon.s$^{-1}$) & 0.21 & 0.6 \\
 \hline
 Photon flux min (photon.s$^{-1}$) & 0.06 & 0.2 \\
 \hline 
 \end{tabular}
 \\[0.3cm]
 \begin{tabular}{|c|c|}
 \hline
 Photon flux ratio & L2/A11 \\
 \hline
 3 Dec 2015 & 3.1 \\
 \hline
 Theoretical ratio & 3 \\
 \hline
 \end{tabular}
 
 \end{table*}

 An effort was made to successively observe A11 and L2 when they were both in the dark. The photon flux ratio between L2 and A11 is supported by 13 measurement series and more than 3500s of cumulative observation time, which give a high degree of confidence in the results. For IR wavelengths, L2 is three times stronger than A11. The discrepancy with the theoretical estimation is only 2\% for the L2/A11 photon flux ratio. 
 
 Photon flux ratios were calculated for consecutive observations during the same night. Using these, the flux ratios over the 9 months of observations were computed. The results from the statistical analysis are summarised in Table \ref{table:3}. The photon flux on A15 is three times stronger that on A11 (or A14), as expected due to the different number of corner cubes between A11 (or A14) and A15 arrays. L1 $\&$ L2 performs similarly to A15. The statistical analysis realised by distinguishing the nights where the reflectors were lit or in the dark gives similar results which are summarised in the Table \ref{table:4}. It is very interesting to note that we obtain a good agreement with the theoretical ratios estimated above, because the concordance is better than 1\% ([(3.1-3)/3 + (3.0-3)/3 + (1-1)/1 + (1-1)/1] / 4  $\sim$ 0.008).
 
 %-------------------------------------------------------------
  %                                             Simple A&A Table
  %-------------------------------------------------------------
  %
  \newcolumntype{C}[1]{>{\centering\arraybackslash }b{#1}}
  \begin{table*}
  \caption{Photon flux ratios between the different reflectors computed over 9 months of observations.}             % title of Table
  \label{table:3}      % is used to refer this table in the text
  \centering                          % used for centering table
  \begin{tabular*}{16.2cm}[]{|C{6cm}|C{2cm}|C{2cm}|C{2cm}|C{2cm}|}
  \hline
  Photon flux ratio & A15/A11 & A15/A14 & A15/L2 & A15/L1 \\
  \hline
  Mean & 3.1 & 3.1 & 1 & 1 \\
  \hline
  Number of nights & 18 & 14 & 12 & 7 \\
  \hline
  Standard deviation & 0.9 & 1.2 & 0.2 & 0.4 \\
  \hline
  Cumulative photon number & 3740 & 3191 & 4255 & 1339 \\
  \hline
  Cumulative observation duration (s) & 22402 & 18610 & 15250 & 8427 \\
  \hline
  \end{tabular*}
  \\[0.1cm]
  \begin{tabular*}{16.2cm}[]{|C{6cm}|C{2cm}|C{2cm}|C{2cm}|C{2cm}|}
      \hline
     Mean for the two reflectors in the dark & 3 & 3.3 & 1 & 1 \\
     \hline
     Number of nights & 3 & 3 & 6 & 6 \\
     \hline
  \end{tabular*}
    \\[0.1cm]
   \begin{tabular*}{16.2cm}[]{|C{6cm}|C{2cm}|C{2cm}|C{2cm}|C{2cm}|}
          \hline
         Mean for the two reflectors in the light & 3.1 & 3.1 & &  \\
         \hline
         Number of nights & 9 & 11 & & \\
         \hline  
     
   \end{tabular*}
   \end{table*}
   %
 
  %
  %-------------------------------------------------------------
  %                                             Simple A&A Table
  %-------------------------------------------------------------
  %
  \begin{table*}
  \caption{Comparison of the different lunar retro-reflectors' efficiency in IR}             % title of Table
  \label{table:4}      % is used to refer this table in the text
  \centering                          % used for centering table
  \begin{tabular}{|c|c|}        % centered columns (4 columns)
  \hline                 % inserts double horizontal lines
  \multicolumn{2}{|c|}{Photon flux ratio over the different lunar reflectors in IR} \\
  \hline
  Theoretical estimation & Grasse measurements \\
  \hline
  Lunokhod arrays = 3 x A11 \& A14 arrays & L2 array = 3.1 x A11 \\
  \hline 
  \multirow{2}*{Lunokhod arrays = 1 x A15 array} & L2 array = 1 x A15 array \\
  \cline{2-2}  & L1 array = 1 x A15 array \\
  \hline  
  \multirow{2}*{A15 array = 3 x A11 \& A14 arrays} & A15 array = 3.1 x A11 array \\
  \cline{2-2}  & A15 array = 3.1 x A14 array \\
  \hline  
  \end{tabular}
  \end{table*}
 \subsubsection{Comparison with results obtained with the green laser link}

 %
   %-------------------------------------------------------------
   %                                             Simple A&A Table
   %-------------------------------------------------------------
   %
%   \newcolumntype{C}[1]{>{\centering\arraybackslash }b{#1}}
   \begin{table*}
   \caption{Comparison of the different lunar retro-reflectors efficiency in green from \citep{Murphy2011}}             % title of Table
   \label{table:5}      % is used to refer this table in the text
   \centering                          % used for centering table
   \begin{tabular}{|c|c|}        % centered columns (4 columns)
   \hline                 % inserts double horizontal lines
   \multicolumn{2}{|c|}{Photon flux ratio over the different lunar reflectors in Green} \\
   \hline
   Theoretical estimation & APOLLO measurements \\
   \hline
   Lunokhod arrays = 1.8 x A11 \& A14 arrays & L1 array = 1 x A11 \& A14 arrays \\
   \hline 
   \multirow{2}*{Lunokhod arrays = 0.6 x A15 array} & L1 array = 0.3 x A15 array \\
   \cline{2-2}  & L2 array = 0.06 x A15 array \\
   \hline  
   L1 array = 1 x L2 array  & L1 array = 6 x L2 array \\
   \hline  
   \end{tabular}
   \end{table*}
 These results in IR can be compared to those published in \citep{Murphy2011} for green laser link, and are summarised in the Table \ref{table:5}. At the Grasse station, we find similar observations for green wavelengths. The mean concordance between the experimental ratio and the theoretical ratio is at the 171\% level (because [ (1.8-1)/1.8 + (0.6-0.3)/0.6 + (0.6-0.06)/0.6 + (6-1)/1 ] /4  $\sim$ 1.71 ]). However, the question remains as to why the estimate does not fit with APOLLO $\&$ Grasse observations at green wavelengths.
 
 \subsection{Comparison of green and infrared links}
 In this subsection, we compared the optical links at the wavelengths of 532 nm and 1064 nm. The number of detected photons $n_{pe}$ is given by the link budget equation \citep{Degnan1993}:
 \begin{equation}
 n_{pe} = \eta_q \cdot (E_t \cdot \frac{\lambda}{hc}) \cdot \eta_t \cdot G_t \cdot \sigma \cdot (\frac{1}{4 \pi R^2})^2 \cdot A_r \cdot \eta_r \cdot T_a^2 \cdot {T_c}^2, \\
 \end{equation}
 
 where $\eta_q$ is the detector quantum efficiency, $E_t$ is the laser pulse energy, $\lambda$ the laser wavelength, h is Planck’s constant, c is the velocity of light in vaccum, $\eta_t$ is the transmit optics efficiency, $G_t$ is the transmitter gain, $\sigma$ is the satellite optical cross section, $R$ is the slant range to the target, $A_r$ is the effective area of the telescope receive aperture, $\eta_r$ is the efficiency of the receive optics, $T_a$ is the one-way atmospheric transmission, and $T_c$ is the one-way transmissivity of cirrus clouds.
 In the following subsections, we highlight the wavelength dependence of the link budget equation. Firstly, we estimate the gain obtained on the signal by changing the working wavelength from green to IR. Secondly, we present the results obtained on A15 for simultaneous measurements in green and in IR. 
 
 \subsubsection{Theoretical estimation}
 \paragraph{Laser.}
 A mode-locked pulsed Nd:YAG laser is used at Grasse for LLR. At the output of the last amplifier we observe a temporally Gaussian pulse at 1064 nm with 150 ps FWHM and an energy of 300 mJ. The repetition rate of the laser is 10 Hz. To generate green pulses, we used a lithium triborate (LBO) crystal for the second harmonic generation (SHG) that has an efficiency ($\eta_{SHG}$) of approximately 65\%. The relative gain between IR and green, relating to the second term of Eq. 3, is defined as the ratio of the photon number at each wavelength:. 
 \begin{equation}
  G_{laser} (\frac{IR}{green}) = \frac{nph_{1064}}{nph_{532}} = \frac{E_t \cdot \frac{\lambda_{1064}}{hc}}{E_t \cdot \frac{\lambda_{532}}{hc} \cdot \eta_{SHG}} \approx 3. \\
 \end{equation}
 The use of the 1064 nm instead of the 532 nm therefore procures three times more photons.
 
 \paragraph{Transmit and receive optics.}
 The laser pulse is then brought up to the telescope output via seven mirrors. Two kinds of mirrors are used. The first are dielectric mirrors that have the same level of reflectivity in green and in IR (98\% from the datasheet). The second, used for the last three mirrors, are metallic mirrors with an aluminium treatment  that have a reflectivity of 95.8\%  for 532 nm wavelength and 94.8\% for 1064 nm wavelength (from the datasheet). At the receiver, the laser echoes are reflected a second time by these three metallic mirrors before being sent to the two detectors. We can estimate an instrumental transmission efficiency ($\eta_t$ in Eq. 3) in green as 0.98$^7$ x 0.958$^3$=76.3\% and in IR as 0.98$^7$ x 0.948$^3$=74\%. The instrumental reception efficiency ($\eta_r$ in Eq. 3) is estimated in green as 0.958$^3$=88\% and in IR as 0.948$^3$=85.2\%. The relative gain between IR and green concerning the transmit and receive efficiency is defined as
 \begin{equation}
   G_{optics} (\frac{IR}{green}) = \frac{\eta_t^{1064} \cdot \eta_r^{1064}}{\eta_t^{532} \cdot \eta_r^{532}} = \frac{0.74 \cdot 0.852}{0.763 \cdot 0.88} = 0.939 \\
 \end{equation}
 in the current Grasse configuration. 
 
 \paragraph{Beam divergence and atmospheric transmission.}
 At Grasse, the whole aperture of the primary mirror is used to transmit both the green and the IR laser pulses. The laser beam is centrally obscured by the secondary mirror (Cassegrain configuration). This obscuration is the same for the two wavelengths and is approximately 0.2. At the telescope output, the energy distribution is uniform over the whole aperture. In far field, this distribution can be approximated by a Gaussian distribution with a beam waist $\omega_0$ equal to the radius of the primary mirror $D_{tel}$ /2. After a propagation in vacuum on a distance R, the radius of a Gaussian beam will be equal to:
 \begin{equation}
 \omega = \frac{R \lambda}{\pi \omega_0} = \frac{2R \lambda}{\pi D_{tel}} \\
 \end{equation}
 
 Therefore at first approximation we estimated that the IR beam width will be two times larger than the green beam width, at the same distance $R$ and for the same initial beam waist $\omega_0$. In this situation, the energy of the IR beam will be more spatially diluted than the energy of the green beam. However, due to atmospheric turbulence outside, the limit of diffraction is not determined by the radius of the primary mirror but by the Fried parameter $r_0$, which depends on local atmospheric conditions. Moreover, $r_0$ is a function of the wavelength \citep{fried1966optical,hardy1998adaptive} and changes at a rate of $\lambda^{6/5}$ resulting in 
 \begin{equation}
  r_0^{1064}=2^{6/5} \cdot r_0^{531} \\
 \end {equation}
 
 Therefore, after a propagation in the atmosphere through a distance $R$, the radius of the Gaussian beams are
 \begin{eqnarray}
 \omega_{532}(R) & = & \frac{2R \lambda_{532}}{\pi r_0^{532}}, \\
  \nonumber \\[0.1cm]
  \omega_{1064}(R) & = & \frac{2R \lambda_{1064}}{\pi r_0^{1064}} \nonumber \\
   & = & \frac{2}{2^{6/5}} \cdot \frac{2R \lambda_{532}}{\pi r_0^{532}} \nonumber \\
   & = & \frac{2}{2^{6/5}} \cdot \omega_{532}(R) \\
  \nonumber 
  \end{eqnarray}
  
 Due to the peak intensity of a Gaussian beam being inversely proportional to the area covered, the relative gain between IR and green concerning the divergence is defined as
 \begin{equation}
    G_{divergence} (\frac{IR}{green}) = \frac{\omega_{532}^2(R)}{\omega_{1064}^2(R)} = (\frac{2^{6/5}}{2})^2 \approx 1.3 \\
 \end{equation}
 
 We can also estimate the difference in atmospheric transmission between the green and the IR beam at sea level \citep{Degnan1993}. These data are summarised in Table \ref{table:6}. Due to the impact of the atmosphere, the gain ratio between IR and green (IR/green) is approximately (1.2)$^{2}$=1.44 at high elevation angles and approximately (1.9)$^{2}$=3.61 at low elevation angles. If we assume that the detection efficiency of the green instrumentation is similar to the IR detection efficiency, we obtain the results summarised in Table \ref{table:7}.  The expected gains presented in Table \ref{table:7} cannot be directly confirmed experimentally. They have been calculated for an optimised laser station working in either green or infrared. In practice we cannot quickly pass from one wavelength to another with all the settings optimised. In the following section, we present results obtained with a laser station set in green that has simultaneous green and IR links.
 
%                                             Simple A&A Table
 %-------------------------------------------------------------
 %
 \newcolumntype{C}[1]{>{\centering\arraybackslash }b{#1}}
 \begin{table}
 \caption{Atmospheric transmission at 532 nm and 1064 nm for different elevation angles. Data are taken from \citep{Degnan1993} and have been calculated for an extremely clear atmosphere.}             % title of Table
 \label{table:6}      % is used to refer this table in the text
 \centering                          % used for centering table
 \begin{tabular*}{8.85cm}[]{|C{1.1cm}|C{2cm}|C{2cm}|C{2cm}|}       % centered columns (4 columns)
 \hline                 % inserts double horizontal lines
 Elevation Angle ($^\circ$) & Atmospheric Transmission at 532 nm (\%) & Atmospheric Transmission at 1064 nm (\%) & IR/green atmospheric transmission ratio \\
 \hline 20 & 50 & 95 & 1.9 \\
 \hline 40 & 75 & 99 & 1.32 \\
 \hline 90 & 81 & 100 & 1.23 \\
 \hline
 \end{tabular*}
 \end{table}
 % 
 
  %
   %-------------------------------------------------------------
   %                                             Simple A&A Table
   %-------------------------------------------------------------
   %
   \begin{table*}
   \caption{Expected relative gain between IR and green on the signal received}             % title of Table
   \label{table:7}      % is used to refer this table in the text
   \centering                          % used for centering table
   \begin{tabular}{|c|c|c|c|c|}        % centered columns (4 columns)
   \hline                 % inserts double horizontal lines
   \multicolumn{5}{|c|}{Expected gain in IR compared to green link} \\
   \hline
   Elevation angle & \multicolumn{2}{c|}{20$^\circ$} & \multicolumn{2}{c|}{40$^\circ$} \\
   \hline
   Retro-reflector & A11/A14/A15 & L1/L2 & A11/A14/A15 & L1/L2 \\
   \hline 
   Laser & \multicolumn{2}{c|}{3} & \multicolumn{2}{c|}{3} \\
   \hline  
   Transmit and receive optics & \multicolumn{2}{c|}{0.939} & \multicolumn{2}{c|}{0.939} \\
   \hline
   Divergence & \multicolumn{2}{c|}{1.3} & \multicolumn{2}{c|}{1.3} \\
   \hline
   Atmospheric transmission & \multicolumn{2}{c|}{(1.9)$^2$} & \multicolumn{2}{c|}{(1.32)$^2$} \\
   \hline
   Velocity aberration & 1.28 & 2.14 & 1.28 & 2.14 \\
   \hline
   Total gain IR/green & 16.9 & 28.3 & 8.2 & 13.7 \\
   \hline
   \end{tabular}
   \end{table*}

 \subsubsection{Comparison of simultaneous green and infrared links on A15}
 Strong variations of the atmospheric transmission during the day make it difficult to compare green and IR LLR measurements taken at different epochs. We chose to exploit measurements taken during the same periods of ten minutes on A15. Each detection channel has its own event timer. The pulse energy has been set by changing the orientation of the LBO crystal to obtain the same number of photons in green and in infrared at the output (200 mJ and 100 mJ respectively. In these conditions a gain of a factor three  related to the laser in Table \ref{table:7} becomes equal to one). In the current configuration, we cannot have green and IR beams of equivalent sizes with divergence optimised for both at the same time. A set up optimised for green wavelengths leads to an IR/green transmission ratio of 0.6. Finally, taking into account telescope transmission, reception and divergence leads to an IR/green transmission ratio of 0.73. With these parameters, we estimate a total gain of IR with respect to green of 3.38 at an elevation angle of 20$^\circ$ and 1.62 at an elevation angle of 40$^\circ$. 
  %----------------------------------------------------------------- one column figure
        \begin{figure}
        \centering
        \includegraphics[width=\hsize]{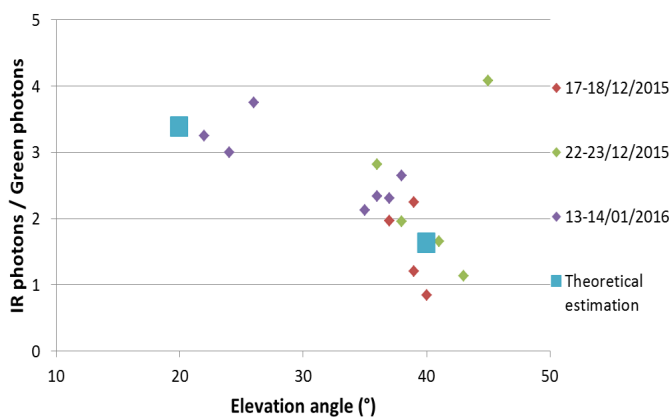}
           \caption{Ratio of the photon flux in IR and in green for different elevation angles.}\label{fig:4}
        \end{figure}
     %-----------------------------------------------------------------

 We experimentally obtain a mean value of 2.1 for observations performed at an elevation angle of 40$^\circ$ +/- 5$^\circ$, with a quite large  standard deviation value of 0.8 that reflects the strong fluctuations of the atmospheric transmission over time. The mean discrepancy with the theoretical estimation is 30\%. 
 
 \subsection{Statistical results over nine months}
\begin{figure}
         \centering
         \includegraphics[width=\hsize]{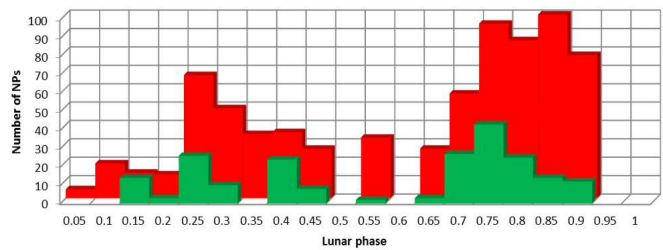}
            \caption{Comparison of green and infrared NPs obtained from January to October 2015 on all the lunar retro-reflectors versus the lunar phase. New Moon matches lunar phase 0.0 and 1.0. Full Moon matches lunar phase 0.5.}\label{fig:5}
         \end{figure}

 We report here our first results obtained over a period of nine months, from January to October 2015. During this period we fixed the instrumental configuration. Simultaneous observations at both wavelengths were preferred, but the SHG was removed when no data was obtained after several attempts. Figure \ref{fig:5} shows the NPs obtained during this period in green and in IR versus the lunar phase. 
 
  %----------------------------------------------------------------- one column figure
      %-----------------------------------------------------------------
 
\begin{figure}
      \centering
      \includegraphics[width=11cm]{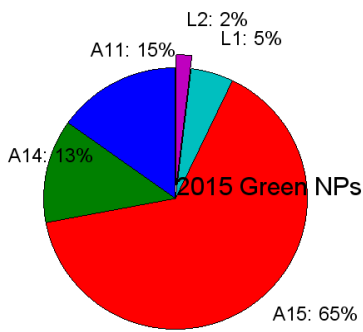}
      \includegraphics[width=11cm]{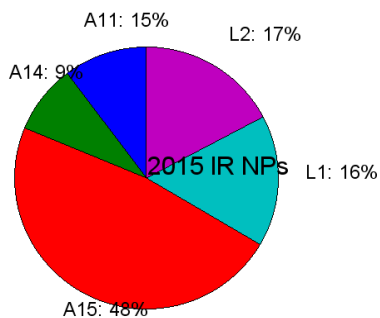}
     \caption{Comparison of the Grasse NPs distribution on the different lunar retro-reflectors in infrared and in green for the first nine months 2015.}\label{fig:6}
       \end{figure}

 Even if a decrease in the number of LLR NPs obtained during new and full Moon periods is still visible, we now have at least obtained observations during these phases. Infrared NPs are systematically more numerous than green ones, for example 741 and 211 NPs have been obtained in IR and in green, respectively. Although midday observation is impossible for green detection, it is possible for those IR detections with  best signal to noise ratio, thus increasing the observability duration. Indeed, observations in green during the year 2014 represent 135 hours whereas observations in IR for the January to October 2015 represent 305 hours. Thus, the observability duration has been multiplied by two for IR. We also obtained interesting results when looking at the observation distribution on the different reflectors (see Fig. \ref{fig:6}). The comparison of the green NPs distribution by target in 2014 and 2015 are quite similar and have a high A15 predominance. This indicates that the inhomogeneity of LLR observations on the different targets does not change during years with green link. On the contrary, IR link obtains a better homogeneity in the targets followed. Even if A15 remains the most observed target, we have obtained very strong links on the Lunokhod arrays in IR. 
 
  %-------------------------------------- Two column figure (place early!)

\begin{figure}
          \centering
          \includegraphics[width=\hsize]{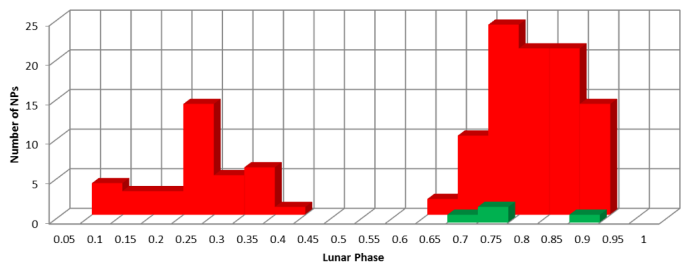}
             \caption{Comparison of green and infrared NPs obtained from January to October 2015 on L2 versus the lunar phase. New Moon matches lunar phase 0.0 and 1.0. Full Moon matches lunar phase 0.5.}\label{fig:7}
          \end{figure}

\begin{figure*}
           \centering
           \includegraphics[width=\hsize]{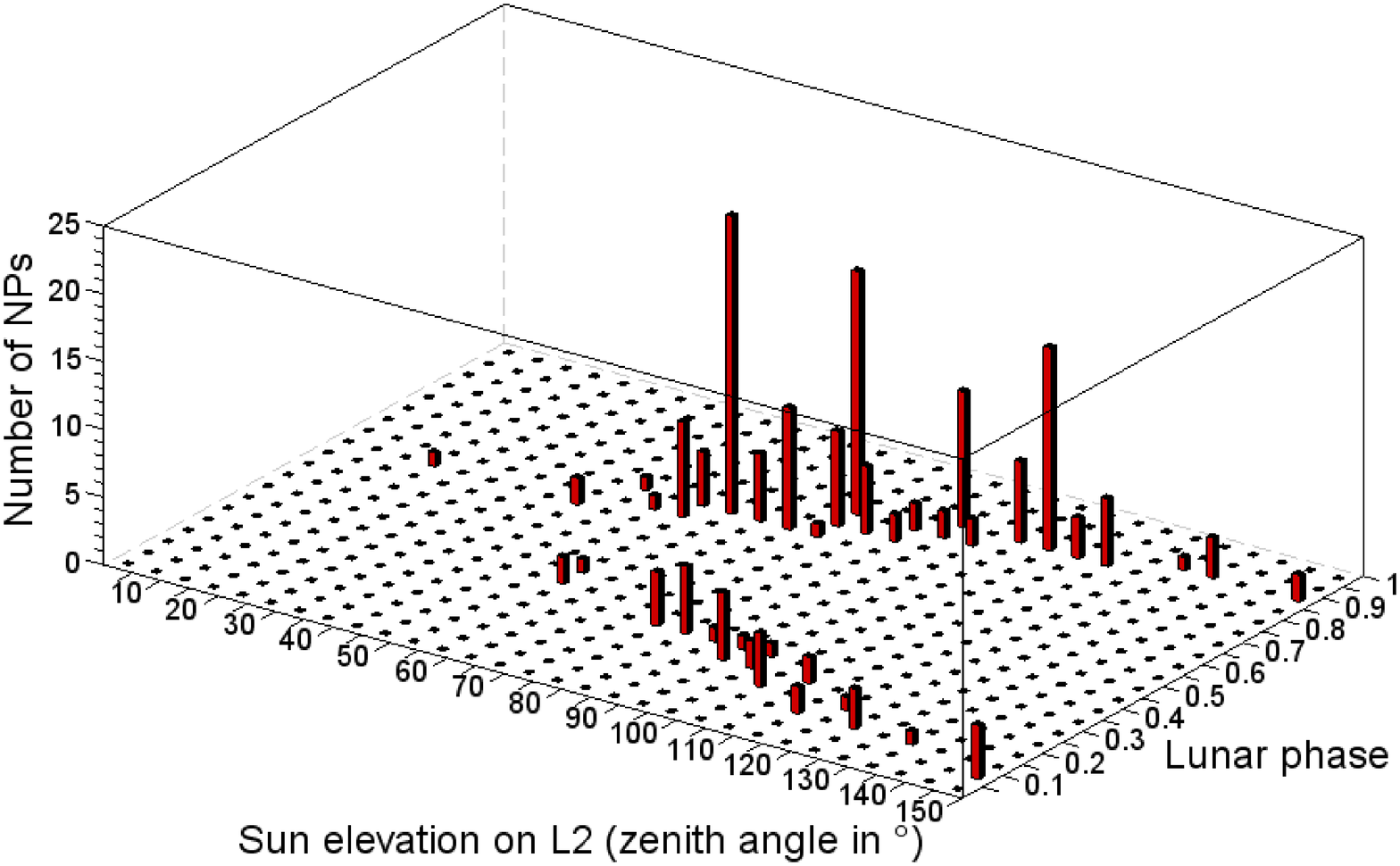}
              \caption{Distribution of the NPs on L2 in IR versus the lunar phase and the Sun elevation on the reflector. New Moon matches lunar phase 0.0 and 1.0. Full Moon matches lunar phase 0.5.}\label{fig:8}
           \end{figure*}

\begin{figure*}
            \centering
            \includegraphics[width=\hsize]{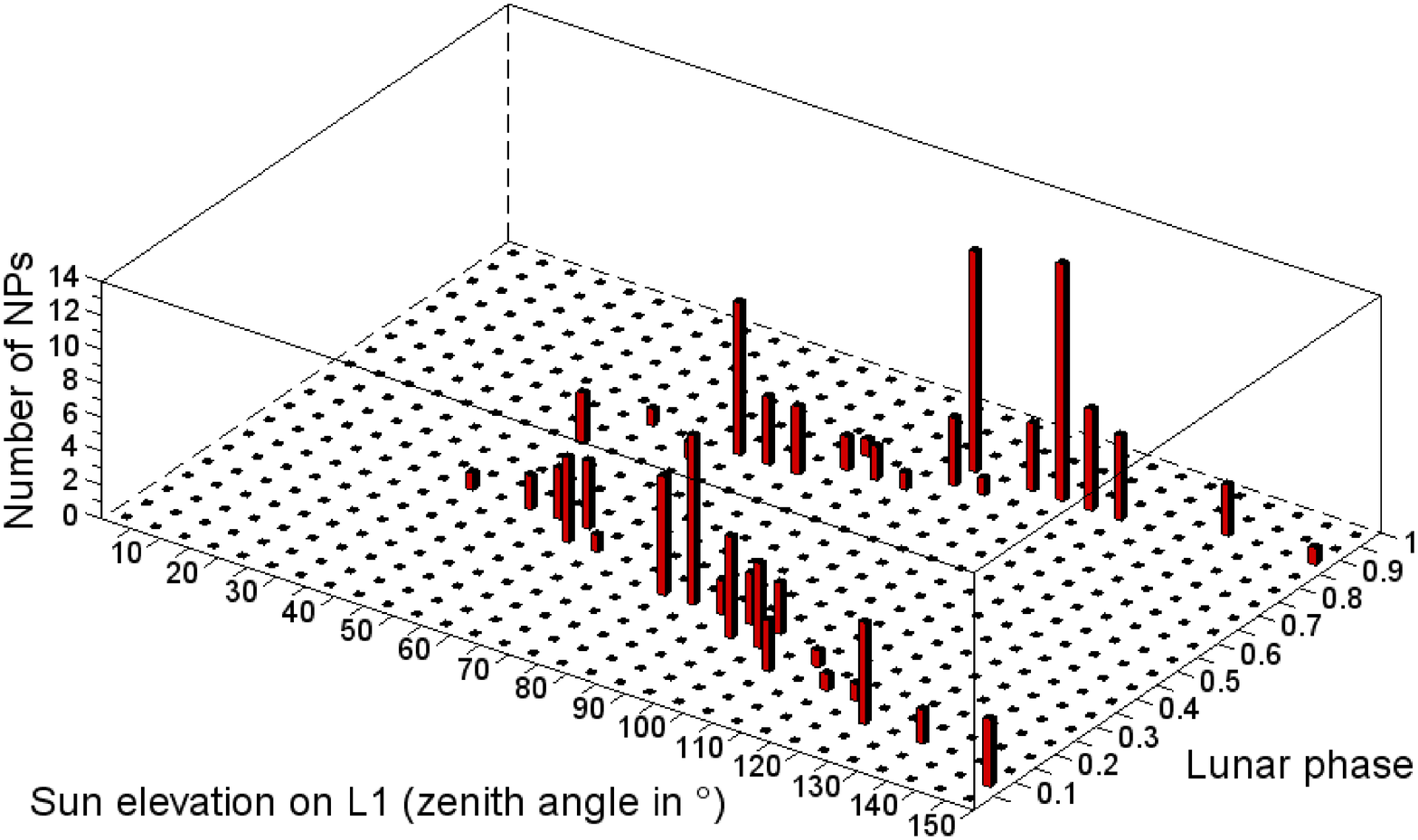}
               \caption{Distribution of the NPs on L1 in IR versus the lunar phase and the Sun elevation on the reflector. New Moon matches lunar phase 0.0 and 1.0. Full Moon matches lunar phase 0.5.}\label{fig:9}
            \end{figure*}

\begin{figure}
             \centering
             \includegraphics[width=\hsize]{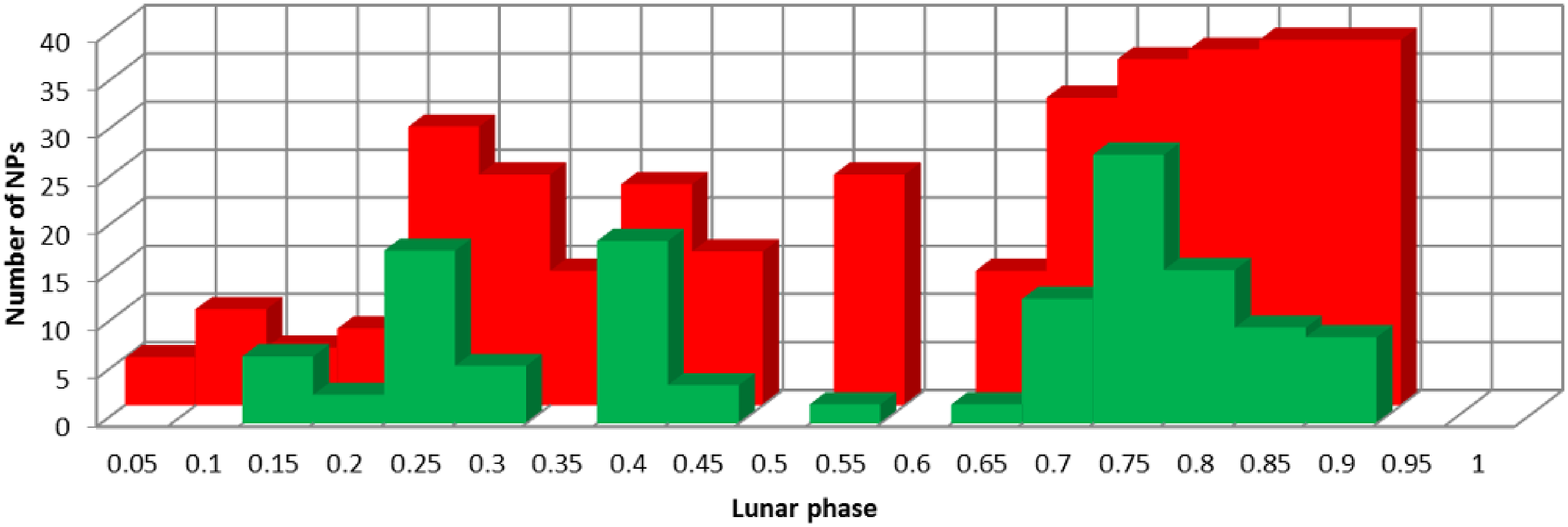}
                \caption{Comparison of green and infrared normal points obtained from January to October 2015 on A15 versus the lunar phase. New Moon matches lunar phase 0.0 and 1.0. Full Moon matches lunar phase 0.5.}\label{fig:10}
             \end{figure}

 The distribution of numbers of NPs versus the lunar phases for each array is also very informative. In Fig. \ref{fig:7}, the gain of IR on green is evident on Lunokhod. Lunokhod arrays have not been designed to work when they are illuminated by the sunlight. That is why Lunokhod observations are impossible at full Moon, whatever the wavelength. The 3D representation in Fig. \ref{fig:8} and Fig. \ref{fig:9}, shows the impact of the sun elevation on the reflector. In IR, we obtained NPs on L2 for a solar elevation on the array of 20$^\circ$. Considering data published in \citep{Fournet1972}, our measurements indicate that we are able to obtain IR echoes despite an efficiency factor reduced to 0.045 (Table \ref{table:1}). These measurements indirectly confirm the expected gains summarised in Table \ref{table:7}. As expected the gain is smaller for A15 (Fig. \ref{fig:10}) but the IR contribution is larger at the new and full Moon periods.  
 
 %----------------------------------------------------------------- one column figure
       %-----------------------------------------------------------------
       
 %----------------------------------------------------------------- one column figure
        %-----------------------------------------------------------------
 %----------------------------------------------------------------- one column figure
         %-----------------------------------------------------------------
  %----------------------------------------------------------------- one column figure
          %-----------------------------------------------------------------
 
 IR detection has enabled an increase in the number of retro-reflectors followed during the same lunar observation. In 2014 with green link, we had just one night with NPs on all the lunar retro-reflectors. In 2015 with IR link, this number increased to twenty.  
 
 %                                             Simple A&A Table
  %-------------------------------------------------------------
  %
  \newcolumntype{C}[1]{>{\centering\arraybackslash }b{#1}}
  \begin{table}
  \caption{Number of nights (inside one lunar cycle) on which different lunar reflectors have been successfully observed.}             % title of Table
  \label{table:8}      % is used to refer this table in the text
  \centering                          % used for centering table
  \begin{tabular*}{8.32cm}[]{|C{3cm}|C{2cm}|C{2cm}|}       % centered columns (4 columns)
  \hline                 % inserts double horizontal lines
  Number of different retro-reflectors followed during the night & Green LLR 2014 night number & IR LLR 2015 night number \\
  \hline 5 & 1 & 20 \\
  \hline 4 & 11 & 8 \\
  \hline 3 & 14 & 18 \\
  \hline
  \end{tabular*}
  \end{table}

 \subsection{Normal point precision and statistical centroid uncertainty}
 The precision of LLR NPs is the result of combining all the different uncertainty sources involved in LLR measurements. In \citep{Samain1998}, all those different uncertainty sources are described. Two main sources can be distinguished, namely the instrumental sources and the retro-reflector array orientation. Because the return detection operates in a single photon mode, an orientation difference between the normal axis of a retro-reflector array and the axis defined by the direction (retroreflector array, telescope) will introduce a dispersion in the measurements. This orientation difference depends on the lunar libration, and the initial orientation of the panel as compared to the mean orientation of the Earth centre as seen from the Moon. Therefore, the NP precision depends strongly on the libration. The correlation between the theoretical and the observed precisions on A15 versus the lunar libration can be found in \citep{Samain1998}. The libration effect on the A15 orientation lies in an interval from 0 to 350 ps, which is 0 to 150 ps for other retro-reflectors. During an LLR observation we can consider the retro-reflector orientation as fixed. The single-shot precision of the station obtained by calibration ranging gives the combined precision of all the instrumental error sources, which Fig. \ref{fig:3} shows are equal to 74 ps RMS for the green channel and 101 ps RMS for the IR channel. The time stability of the green and IR calibrations shows a white-phase noise during a typical LLR observation time. During an LLR observation where the retro-reflector dispersion is equal to zero, we should have a NP precision close to the single-shot precision of the calibration because the atmosphere effect on the path fluctuation is in the range of units of picoseconds \citep{kral2005optical}. The NP statistical centroid uncertainty $\sigma_r$ is given by the Eq. 11, with $N_{count}$ the number of measurements during the LLR observation: 
 
  \begin{equation}
   \sigma_{r} = (\frac{\sigma_{single\,shot\,precision}}{\sqrt{N_{count}}}). \\
  \end{equation} 
 
 The $\sigma_r$ value in picoseconds can be expressed as a one-way range uncertainty $\sigma_{mm}$ in mm with $\sigma_{mm} = \sigma_r \cdot c / 2 $. Because the single-shot precision of the IR channel is larger than the green one, more measurements are needed in IR to obtain the same NP statistical centroid uncertainty. Using Eq. 11 and the single-shot precision of each detection channel, we calculate that the number of IR counts should be equal to two ($\frac{101^2}{74^2} \approx 2$) times the number of green counts. Moreover, 230 counts and 124 counts are respectively needed in IR and in green to reach a millimetric one-way range uncertainty. 
 
\begin{figure*}
       \centering
       \includegraphics[width=\hsize]{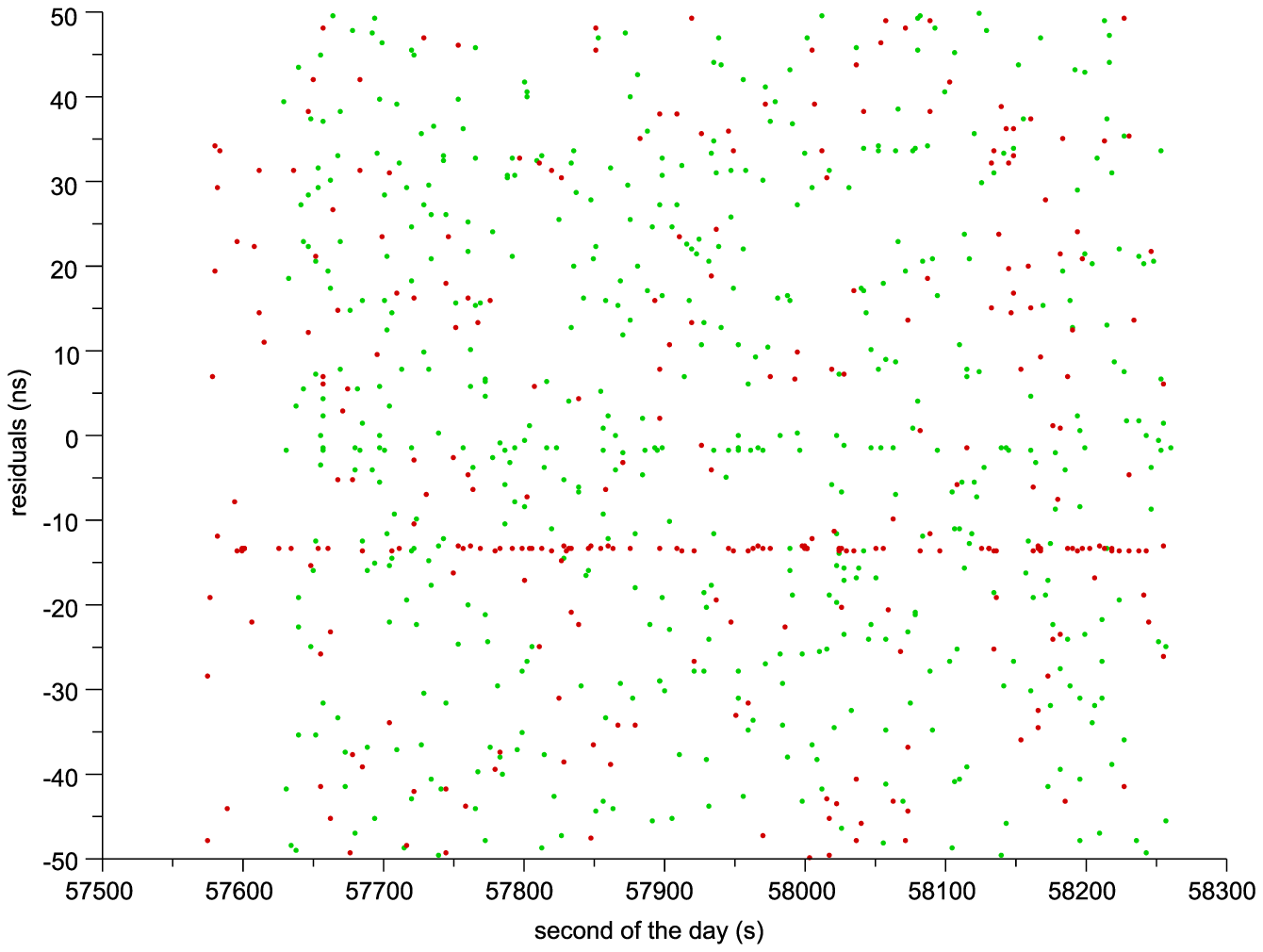}
       \includegraphics[width=\hsize]{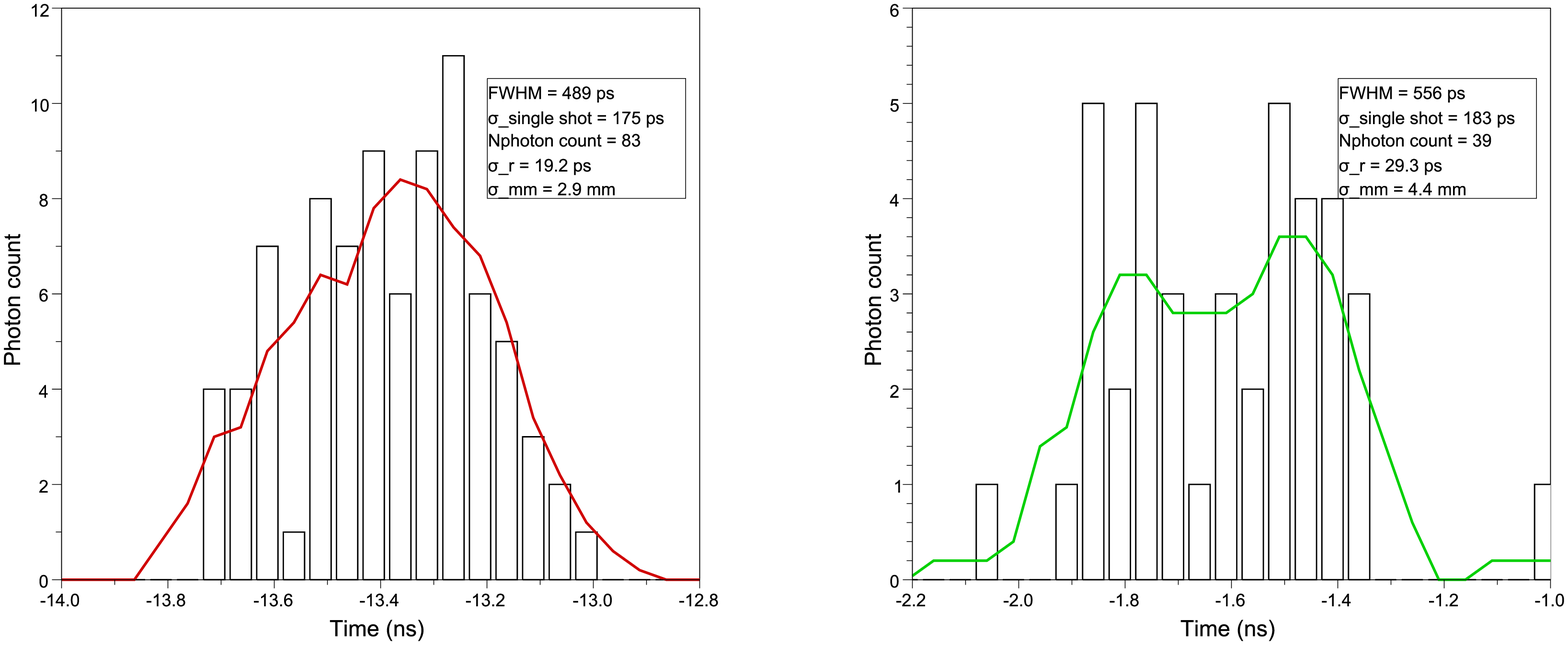}
      \caption{Two-colour LLR measurement on A15, performed on 13 Jan 2016 at 15H59 UTC. The upper part shows the green and IR counts as obtained in realtime. The lunar returns are the two tracks which are clearly separated from the noise. The scale of the residual is arbitrary. The IR counts are offset from the green counts to improve their visibility. The lower part shows the histogram of the IR and green counts. The histograms are created using 50 ps bins. The traces in bold are a five-point moving average. The two peaks present for the green are due to a data-binning effect.}\label{fig:11}
        \end{figure*}

 Figure \ref{fig:11} presents an example of LLR measurements on A15 simultaneously in green and in IR. We see that the noise is smaller for the IR detection channel. The precisions in green and in IR are equivalent. Two times more counts have been obtained in IR, allowing a better one-way range uncertainty to be obtained.
 
 %-------------------------------------- Two column figure (place early!)

\begin{figure}
        \centering
        \includegraphics[width=\hsize]{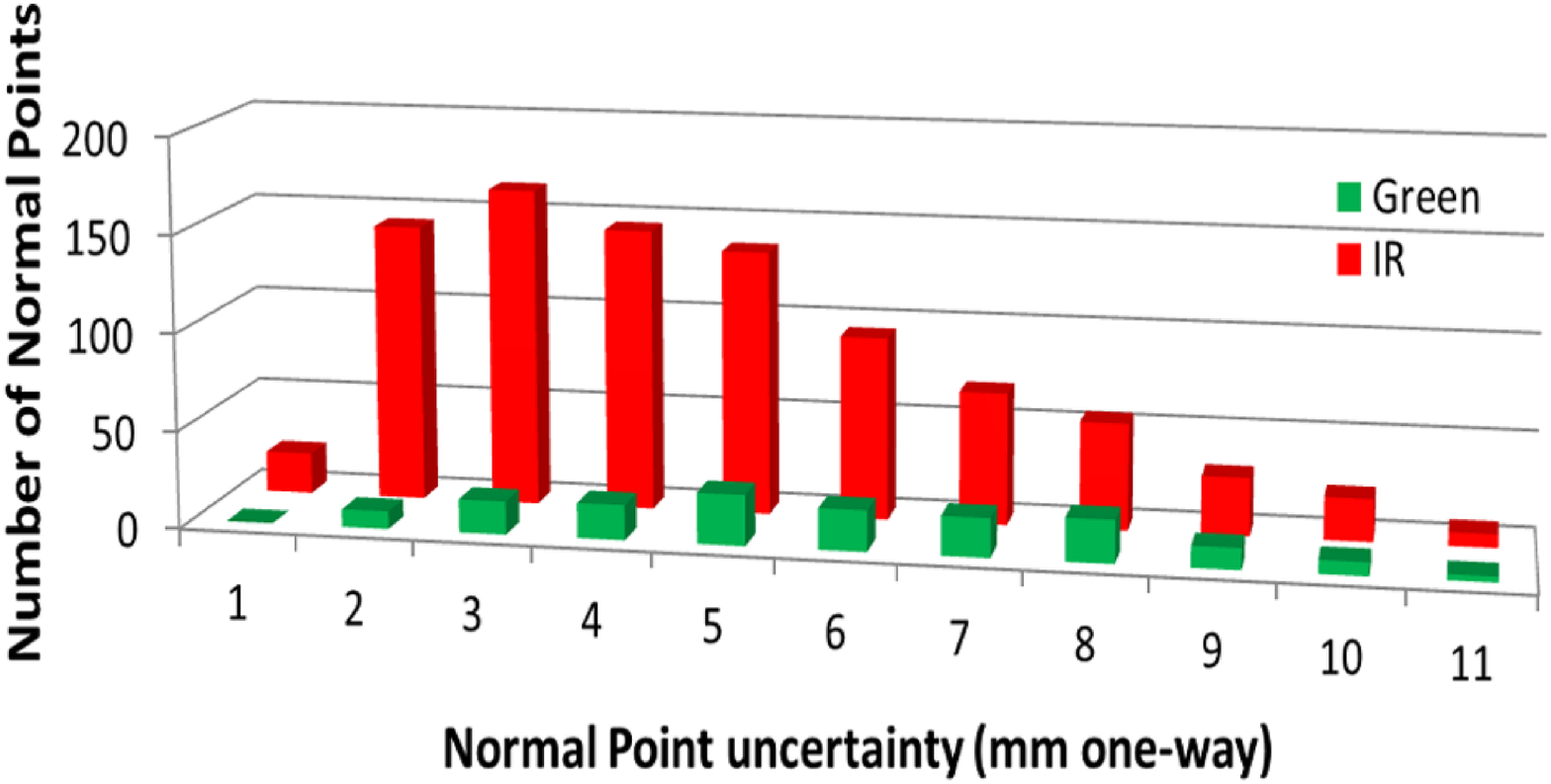}
       \caption{Distribution of the green and IR NPs in 2015 versus their one-way range uncertainties.}\label{fig:12}
         \end{figure}

 Figure \ref{fig:12} presents the distribution of the green and IR NPs in 2015 versus their one-way range uncertainties. The maximum of the IR distribution is at the three millimetre level compared to five millimetre for green. This improvement is explained by two points. Firstly, more counts are obtained in IR during an LLR observation, leading to better one-way range uncertainties. Secondly, more NPs are obtained on the L1 and L2 retro-reflectors that are less dispersive than A15 due to their smaller size. Therefore, IR detection gives more precise NPs than those obtained in green.    
 
  %-------------------------------------- Two column figure (place early!)
\begin{figure}
         \centering
         \includegraphics[width=\hsize]{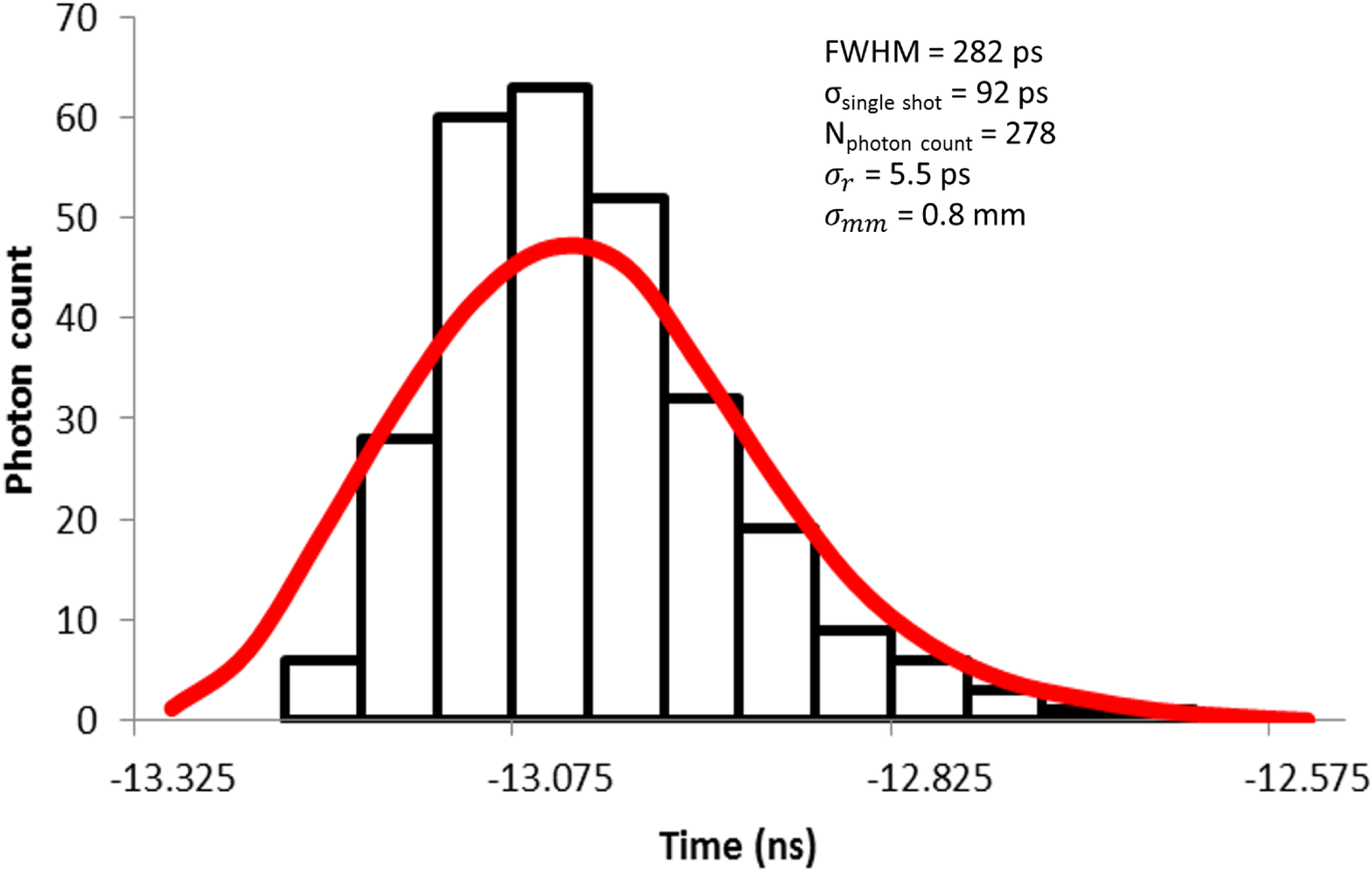}
        \caption{Histogram of LLR measurements on L1 in IR on 15 Jan 2016. The histogram is created using 50 ps bins. The trace in bold is a five-point moving average.}\label{fig:13}
          \end{figure}

 Figure \ref{fig:13} presents the histogram of LLR measurements on L1 in IR on 15 Jan 2016. The sigma of the measurements is in the range of the single-shot precision of the calibration. We obtain a one-way range uncertainty better than 1 millimetre because we collected 278 counts.
 
  %-------------------------------------- Two column figure (place early!)

 From March 2015 to May 2016, we obtained four NPs in IR with sub-millimetric one-way range uncertainty. These results are reported in Table \ref{table:9}. Unlike at the APOLLO station, sub-mm one way range uncertainty data is not routinely available with green detection at Grasse. However, this data is now accessible with IR detection at Grasse. 
  %                                             Simple A&A Table
   %-------------------------------------------------------------
   %
   \newcolumntype{C}[1]{>{\centering\arraybackslash }b{#1}}
   \begin{table*}
   \caption{Normal point with sub-millimetric one-way range uncertainty.}             % title of Table
   \label{table:9}      % is used to refer this table in the text
   \centering                          % used for centering table
   \begin{tabular*}{11.2cm}[]{|C{2cm}|C{1cm}|C{2cm}|C{2cm}|C{2cm}|}       % centered columns (5 columns)
   \hline                 % inserts double horizontal lines
   Date & Array & Number of IR counts & NP sigma (ps) & NP one-way range uncertainty (mm) \\
   \hline 7 Sep 2015 & L2 & 475 & 129 & 0.9 \\
   \hline 4 Dec 2015 & L1 & 297 & 114 & 1 \\
   \hline 15 Jan 2016 & L1 & 278 & 94 & 0.8 \\
   \hline 14 Mar 2016 & L1 & 474 & 114 & 0.8 \\
   \hline
   \end{tabular*}
   \end{table*}
 Table \ref{table:10} summarises the results obtained at Grasse over more than one year in green and in IR. The impact of IR on the NP number, on the count number per NP, and on the one-way range uncertainty are all positive. Because the single-shot precision of the calibration is larger in IR than green, it is not surprising to have an NP sigma that is a little bit larger in IR. Finally, the IR impact is very impressive on L2 meaning that the Grasse station NPs are a little bit more accurate than APOLLO NPs.  
 
  %-------------------------------------------------------------
    %                                             Simple A&A Table
    %-------------------------------------------------------------
    %
    \begin{table*}
    \caption{Statistical results over the 11 Mar 2015 - 17 May 2016 period}             % title of Table
    \label{table:10}      % is used to refer this table in the text
    \centering                          % used for centering table
    \begin{tabular}{|c|c|c|c|c|c|c|c|c|c|}        % centered columns (10 columns)
    \hline                 % inserts double horizontal lines
     & \multicolumn{2}{c|}{Number of} & \multicolumn{2}{c|}{Median of}  & \multicolumn{2}{c|}{Median of} & \multicolumn{2}{c|}{Median of} & Median of \\
     & \multicolumn{2}{c|}{NPs} & \multicolumn{2}{c|}{count number}  & \multicolumn{2}{c|}{NP $\sigma$} & \multicolumn{2}{c|}{NP one-way} & NP one-way \\
     & \multicolumn{2}{c|}{} & \multicolumn{2}{c|}{ per NP}  & \multicolumn{2}{c|}{(ps)} & \multicolumn{2}{c|}{range } & range \\
     & \multicolumn{2}{c|}{} & \multicolumn{2}{c|}{}  & \multicolumn{2}{c|}{} & \multicolumn{2}{c|}{uncertainty} & uncertainty at the \\
     & \multicolumn{2}{c|}{} & \multicolumn{2}{c|}{}  & \multicolumn{2}{c|}{} & \multicolumn{2}{c|}{(mm)} & APOLLO station \\
     & \multicolumn{2}{c|}{} & \multicolumn{2}{c|}{}  & \multicolumn{2}{c|}{} & \multicolumn{2}{c|}{} & \citep{Murphy2012} \\
     & \multicolumn{2}{c|}{} & \multicolumn{2}{c|}{}  & \multicolumn{2}{c|}{} & \multicolumn{2}{c|}{} & (mm) \\
     \hline
    Array & green & IR & green & IR & green & IR & green & IR & green \\
    \hline 
    A11 & 27 & 135 & 25 & 40 & 156 & 163 & 5.2 & 3.9 & 2.4 \\
    \hline 
    A14 & 19 & 97 & 25 & 40 & 163 & 178 & 4.7 & 3.9 & 2.4 \\
    \hline 
    A15 & 137 & 575 & 33 & 53 & 255 & 271 & 6.6 & 5.6 & 1.8 \\
    \hline 
    L1 & 9 & 172 & 15 & 43 & 100 & 142 & 3.4 & 3.2 & 2.7 \\
    \hline 
    L2 & 1 & 188 & 6 & 60 & 165 & 153 & 10.1 & 2.9 & 3.3 \\
    \hline  
    \end{tabular}
    \end{table*}

 \section{Scientific impact}
 IR detection has a significant impact on the number of echoes obtained in LLR and the improved returns obtained on the Lunokhod arrays will have a valuable impact on LLR research. Due to their smaller size, the Lunokhod arrays give NPs with the best precision, such that the NPs standard deviation on L1 $\&$ L2 is two times lower than these obtained on A15. The larger number of observations on the Lunokhod arrays should improve the precision of scientific products obtained as a result of LLR data. Knowledge of the Moon’s orientation is crucial for studying many aspects of gravity and the internal structure of the Moon. Using the five reflectors during the same lunar observation doubles the number of independent constraints on the deformation models that are needed to constrain models of the Moon’s tidal deformations \citep{Williams2015}. The IR detection on the Grasse station now gives us this capability, except around full Moon at which point the Lunokhod arrays are unusable both in green and in IR. The Lunokhod arrays provide sensitivity to both longitudinal and latitudinal librations. This is an advantage over the Apollo reflectors that tend to be sensitive only to longitudinal librations. The fact that L2 responds as well as L1 in IR should help to enforce lunar libration constraints on selenophysical models. The change from green to IR wavelengths could also be beneficial for the test of the equivalence principle (EP). It is currently difficult to clearly extract an EP-violating signal which is synodic, because LLR data obtained with green detection do not uniformly sample the synodic month cycle. Moreover, the full potential of the EP test is not exploited with LLR because at the epochs of possible maximum amplitude there are either no (new Moon) or fewer (full Moon) observations \citep{Hofmann2010}. Infrared detection now gives us the ability to observe during all the lunar phases. In this way, we follow the LLR observation recommendations \citep{Muller1998} and reach this objective. Finally, the implementation of IR detection in other LLR stations, that are either in development or already operational, should lead to an overall improvement. Simulations in \citep{Hofmann2013,viswanathan2016llr} show that the observation of as many reflectors as possible during an LLR session (i.e. using the usual five lunar reflectors) will increase the accuracy of estimated parameters such as position and velocities for lunar orbit and rotation, reflector and station coordinates, some lunar gravity field coefficients, tidal parameters, and the mass of the Earth-Moon system by a factor of between 1.8 to 4.6. Moreover, adding an LLR station in the southern hemisphere would lead to an improvement of about 10\%-15\% for the considered parameters, as demonstrated in \citep{FiengaProc}. An error analysis based on the locations of McDonald and Grasse laser stations shows that the addition of one or more reflectors on Moon's surface would improve the geometrical precision of NP by a factor of 1.5 to nearly 4 at the same level of ranging \citep{Merkowitz2007}. At the time of the publication of \citep{Merkowitz2007}, L1 was lost. The rediscovery of L1 coupled with IR detection would help to reach this level of improvement.
 
%-----------------------------------------------------------------

\section{Conclusions}

  The implementation of IR detection for LLR provides new opportunities for the improvement of scientific products. As expected, LLR in infrared increases the station efficiency by a factor of eight during new and full Moon periods and improves the temporal homogeneity of LLR observations over a synodic month. The best link budget at this wavelength results in a significant increase of the NPs over all the reflectors on the Moon. Our observations are statistically more homogenous over all the targets. A surprising result concerns the L2 array behaviour in IR. The degradation of L2 performances compared to L1 seems to be chromatic because the significant difference observed in green at Grasse and APOLLO is absent in IR. Although this difference remains unexplained, the differential reflectivity at those two separate wavelengths gives new insights to analyse the performances of lunar retro-reflectors. 

\begin{acknowledgements}
     We thank the INSU-CNRS and the CNES for their support. 
\end{acknowledgements}

\bibpunct{(}{)}{;}{a}{}{,} % to follow the A&A style
\bibliographystyle{aa} % style aa.bst
\bibliography{article_lune_v2} % your references Yourfile.bib

\end{document}